\documentclass[%
 reprint,
superscriptaddress,
 amsfonts,amsmath,amssymb,amsthm,
 aps,
prx,
]{revtex4-1}

\usepackage{graphicx}
\usepackage{float}
\usepackage{dcolumn}
\usepackage{bm}
\usepackage{verbatim}
\usepackage{xcolor}

\begin{document}



\title{Ultra-fast real-time quantum random number generator with correlated measurement outcomes and rigorous security certification}

\author{Tobias Gehring}
\email{tobias.gehring@fysik.dtu.dk}
\affiliation{Center for Macroscopic Quantum States (bigQ), Department of Physics, Technical University of Denmark, Fysikvej, 2800 Kgs. Lyngby, Denmark}
\author{Cosmo Lupo}
\affiliation{Department of Computer Science, University of York, York YO10 5GH, United Kingdom}
\affiliation{Department of Physics and Astronomy, University of Sheffield, United Kingdom}
\author{Arne Kordts}
\author{Dino Solar Nikolic}
\author{Nitin Jain}
\author{Tobias Rydberg}
\affiliation{Center for Macroscopic Quantum States (bigQ), Department of Physics, Technical University of Denmark, Fysikvej, 2800 Kgs. Lyngby, Denmark}
\author{Thomas B.\ Pedersen}
\affiliation{Cryptomathic A/S, Jaegergardsgade 118, 8000 Aarhus C, Denmark}
\author{Stefano Pirandola}
\affiliation{Department of Computer Science, University of York, York YO10 5GH, United Kingdom}
\author{Ulrik L.\ Andersen}
\email{ulrik.andersen@fysik.dtu.dk}
\affiliation{Center for Macroscopic Quantum States (bigQ), Department of Physics, Technical University of Denmark, Fysikvej, 2800 Kgs. Lyngby, Denmark}

\date{\today}

\begin{abstract}
Quantum random number generators (QRNGs) promise perfectly unpredictable random numbers. 
However, the security certification of the random numbers in form of a stochastic model often introduces assumptions that are either hardly justified or indeed unnecessary. Two important examples are the restriction of an adversary to the classical regime as well as negligible correlations between consecutive measurement outcomes. Additionally, non-rigorous system characterization opens a security loophole. In this work we experimentally realize a QRNG that does not rely on the aforementioned assumptions and whose stochastic model is established by a rigorous -- metrological -- approach. Based on quadrature measurements of vacuum fluctuations, we demonstrate a real-time random number generation rate of 8 \,GBit/s.
Our security certification approach offers a number of practical benefits and will therefore find widespread applications in quantum random number generators. 
In particular, our generated random numbers are well suited for today's  conventional and quantum cryptographic solutions.
\end{abstract}

\maketitle
%
%
%
Random numbers are ubiquitous in modern society. They are used in numerous applications ranging from cryptography, simulations and gambling, to fundamental tests of physics. For most of these applications, the quality of the random numbers is of utmost importance. If, for instance, cryptographic keys originating from random numbers are predictable, it will have severe consequences for the security of the internet. To ensure the security of cryptographic encryption, the random numbers must be truly random, i.e.\ completely unpredictable to everyone and thus private, and their randomness must be certified by establishing a stochastic model~\cite{Frauchiger2013,Acin2016}.

True unpredictability and privacy of the generated numbers can be attained through a quantum measurement process: By performing a projective measurement on a pure quantum state, and ensuring that the state is not an eigenstate of the measurement projector, the outcome is unpredictable and thus true random numbers can be generated. Moreover, since a pure state cannot be correlated to any other state in the universe, the generated numbers will be private. 

Numerous different types of quantum random number generators (QRNGs) have been devised exploiting the quantum uncertainty in photon counting measurements, phase measurements or quadrature measurements~\cite{Ma2016,Herrero-collantes2017,Pirandola2019}. One particular approach of increasing interest due to its high practicality is the optical quadrature measurements of the vacuum state by means of a simple homodyne detection~\cite{Gabriel2010,Symul2011, Avesani2018}. 
This approach combines simplicity, cost-effectiveness, chip-integrability and extraordinary high generation speed.

Previous QRNG implementations -- independently of the method they are based on -- fall short of one or more of the following critical issues:

Adversaries are often assumed to have restricted power: It is often assumed that they only have access to classical side-information and thus have no quantum capabilities. Recently, this issue has been addressed for a source-independent QRNG~\cite{Marangon2017,Avesani2018}, which however requires a more complex measurement apparatus than simple homodyne detection. 

Furthermore, it has often been assumed that measurements are uncorrelated in time~\cite{Gabriel2010,Furst2010,Symul2011,Xu2012,Haw2015,Nie2015,Shi2016,Abellan2014,Zhang2016,Avesani2018,Huang2019,Zheng2019}, despite the fact that the finite bandwidth of a real detection system introduces correlations. Aliasing in the sampling procedure may minimize correlations as well as suitable post-processing algorithms, however such measures usually throttle the overall rate considerably or remove the correlations only partially.

Most previous implementations did not use a conservative and rigorous approach -- a metrology-grade approach~\cite{Mitchell2015} -- to characterize the parameters of the stochastic model that determines the amount of randomness. A rigorous characterization of the system is however of utmost importance as any parameter uncertainty introduces a non-zero probability for system failure, i.e.\ the probability that the actual device does not follow the stochastic model describing the underlying physical random number generation process. Knowing the failure probability for the system is critical to its certification.

Finally, high-speed (GBit) randomness extraction using an information theoretically secure randomness extractor has only been demonstrated recently~\cite{Zhang2016,Zhang2016a,Huang2019,Zheng2019} and thus many reported QRNGs either achieve only moderate speeds or do not even extract random numbers in real-time.

Here, we report on a QRNG which solves all these aforementioned issues simultaneously for the first time. Using a QRNG based on the quadrature measurement of vacuum fluctuations,  
we 1) compute a lower bound on the extractable randomness against a quantum-enabled adversary,
2) account for correlated samples resulting from the finite bandwidth of the measurement apparatus, and 3) perform a metrology-grade characterization of the measuring homodyne detector system to establish the stochastic model.
Finally, as a result, we produce random numbers in real-time with a rate of 8\,Gbit/s using a Toeplitz randomness extractor on a fast field-programmable-gate-array (FPGA).

\begin{figure*}[ht]
  \includegraphics{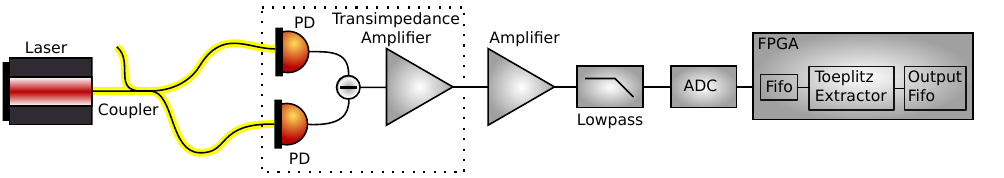}
  \caption{Schematic of the QRNG. A 1.6\,mW 1550\,nm laser beam was split into two by a 3\,dB fiber coupler and detected by a home-made homodyne detector based on a MAR-6 microwave amplifier from Minicircuits and two $120\,\mu m$ InGaAs photo diodes (PD). The output of the detector was amplified with another microwave amplifier, lowpass filtered at 400\,MHz and digitized with a 16\,bit 1 GSample/s analog-to-digital (ADC) converter. The ADC output was read by a Xilinx Kintex UltraScale FPGA. The ADC and FPGA were hosted by a PCI Express card from 4DSP (Abaco). The FPGA was used for real-time randomness extraction based on Toeplitz hashing. Fifo: First-in-first-out buffer.}
  \label{fig:exp}
\end{figure*}

\section{Setting the stage}

A schematic of our QRNG is shown in Fig.~\ref{fig:exp}. An arbitrary quadrature of the vacuum state is measured using a balanced homodyne detector comprising a bright reference beam, a symmetric beam splitter and two photodiodes~\cite{Shapiro1985}. The measurement outcomes ideally are random with a Gaussian distribution associated with the Gaussian Wigner function of the vacuum quantum state~\cite{Weedbrook2012}. The measured distribution, however, contains two additional independent Gaussian noise sources; excess optical noise and electronic noise, thereby contributing two side channels. These must be accounted for in estimating the min-entropy of the source.  

The amount of quantum randomness that can be extracted from the homodyne measurement of vacuum fluctuations is given by the leftover hash lemma~\cite{Renner2005,Tomamichel2011}
\begin{equation}
    \ell \le NH_\text{min}(X|E) - \log\frac{1}{\epsilon_\text{hash}^2}\ .
    \label{eqn:securelength}
\end{equation}
Here $H_\text{min}(X|E)$ is the min-entropy of a single measurement outcome drawn from a random variable $X$ conditioned on the quantum side information $E$ held by an adversary, $N$ is the number of aggregated samples and $\epsilon_\text{hash}$ is the distance between a perfectly uniform random string and the string produced by a randomness extractor. It is therefore clear that we need to find the min-entropy of our practical -- thus imperfect -- realization in order to bound the amount of randomness. We achieve this in a two-step approach: First we theoretically derive a bound for the min-entropy using a realistic security model and express it in terms of experimentally accessible parameters. Second, we experimentally deduce these parameters through a metrology-grade characterization~\cite{Mitchell2015}.
Using such an approach, we find the worst-case 
min-entropy compatible with the confidence intervals of our characterization and calibration measurements, thereby obtaining a string of $\epsilon$-random bits that are trustworthy with the same level of confidence. 


\section{Security analysis}

The security analysis of the QRNG is made under the assumption that the quantum noise is Gaussian and stationary. Therefore, the QRNG follows a device-dependent security model, i.e.\ an adversary cannot change the system after system characterization has been performed. Stationarity is formally expressed by a Wigner function that takes the form of a stationary Gaussian distribution in phase-space. As we are dealing with Gaussian states, the Wigner function is completely characterized by the first and second moments of the field quadratures.

In our analysis we assume that homodyne detection is defined on a single optical mode, which at a given time is characterized by the field quadratures $\hat q$ and $\hat p$. We also make a further assumption that the state is symmetric under a rotation of the quadratures
\begin{align}
\hat q & \to \hat q \cos{\theta} + \hat p \sin{\theta} \, , \\
\hat p & \to \hat p \cos{\theta} - \hat q \sin{\theta} \, ,
\end{align}
for all $\theta \in [0,2\pi]$.

In the following Sections, we first assess the security of a source emitting i.i.d.\ (independent and identically distributed) signals, i.e., a source of infinite bandwidth. We then extend the security analysis to a source with finite bandwidth that emits correlated (non-i.i.d.) signals at different times.

\subsection{IID limit}\label{Sec:iidlimit}

First consider an ideal i.i.d.\ scenario with unlimited bandwidth.
As explained above, we assume that the source emits a Gaussian state of light that is symmetric in phase space. In the i.i.d.\ limit this corresponds to a source of thermal light. We therefore assume a thermal source characterized by a mean photon number per mode of $n$. Each mode is measured independently by homodyne detection.
In other words, our security analysis holds under the assumption that a potential eavesdropper performs a collective Gaussian attack that preserves the above symmetry in the phase space.

For a thermal state $\rho$, the first moments of the field quadratures vanish, and the covariance matrix (CM) is
\begin{align}
V_\mathrm{thermal} 
& = \left( 
\begin{array}{cc}
\langle \hat q^2 \rangle              & \frac{1}{2}\langle \hat q\hat p+\hat p\hat q \rangle \\
\frac{1}{2}\langle \hat p\hat q+\hat q\hat p \rangle & \langle \hat p^2 \rangle
\end{array}
\right) \\
& = \left( 
\begin{array}{cc}
1+2n & 0 \\
0 & 1+2n  
\end{array}
\right) \, ,
\end{align}
where we, as a matter of convention, put the variance of the vacuum equal to $1$. In the equation above we use $\langle\hat{O}\rangle := \text{tr}(\rho\hat{O})$ for operator $\hat{O}$.

For such a state, the output $X$ of ideal homodyne detection is a continuous (real-valued) variable whose probability distribution is 
\begin{align}
p_X(x) = G(x ; 0 , g^2(1+2n)) \, ,
\end{align}
where $g$ is a gain factor and 
\begin{equation}
G(x ; \mu , v^2 ) = \frac{1}{\sqrt{2\pi}v}\,  e^{ - \frac{(x-\mu)^2}{2v^2}}
\end{equation}
denotes a Gaussian in the variable $x$, with mean $\mu$ and variance $v^2$.


To account for quantum side information we assume that a malicious adversary holds a purification of the thermal state emitted by the source. It is well known that the purification of a thermal state can be assumed to be a two-mode squeezed vacuum (TMSV) state without loss of generality (more precisely up to isometries over the environment).
Thereby one optical mode of this TMSV state, characterized by the field quadratures $\hat q_e$ and $\hat p_e$, is controlled by the adversary.
The TMSV state is a Gaussian state with zero mean and CM~\cite{Weedbrook2012}
\begin{widetext}
\begin{align}
V & = 
\left( 
\begin{array}{cccc}
\langle \hat q^2 \rangle              & \frac{1}{2}\langle \hat q\hat p+\hat p\hat q \rangle & \langle \hat q\hat q_e \rangle                     & \langle \hat q\hat p_e \rangle \\
\frac{1}{2}\langle \hat p\hat q+\hat q\hat p \rangle & \langle \hat p^2 \rangle              & \langle \hat p\hat q_e \rangle                     & \langle \hat p\hat p_e \rangle \\
\langle \hat q_e\hat q \rangle             & \langle \hat q_e\hat p \rangle             & \langle \hat q_e^2 \rangle                    & \frac{1}{2}\langle \hat q_e\hat p_e+\hat p_e\hat q_e \rangle \\
\langle \hat p_e\hat q \rangle             & \langle \hat p_e\hat p \rangle             & \frac{1}{2}\langle \hat p_e\hat q_e+\hat q_e\hat p_e \rangle & \langle \hat p_e^2 \rangle 
\end{array}
\right) \\
& = \left( 
\begin{array}{cccc}
1+2n & 0 & 2\sqrt{n(n+1)} & 0 \\
0 & 1+2n & 0 & -2\sqrt{n(n+1)} \\
2\sqrt{n(n+1)} & 0 & 1+2n & 0 \\
0 & -2\sqrt{n(n+1)} & 0 & 1+2n 
\end{array}
\right) \, .
\end{align}
\end{widetext}

The correlations between the outcome $X$ of ideal homodyne detection and the quantum side information held by the eavesdropper are described by the classical-quantum (CQ) state
\begin{equation}\label{CQstate}
\rho_{XE} = \int dx \, p_X(x) |x\rangle \langle x | \otimes \rho_E^x \, ,
\end{equation}
where $|x\rangle$ are orthogonal states used to represent the possible outcomes of homodyne detection, and the integral in Eq.\ (\ref{CQstate}) extends over the real line.
The state $\rho_E^x$ is the conditional state of the eavesdropper for a given measurement output value $x$. Assume, without loss of generality, that the quadrature $\hat q$ is measured. It is then straightforward to compute the first moment of the field quadratures of $\rho_E^x$:
\begin{align}
\left( \begin{array}{c}
\langle \hat q_e \rangle \\
\langle \hat p_e \rangle
\end{array} \right)
=
\left( \begin{array}{c}
\frac{2\sqrt{n(n+1)}}{g(1+2n)} x \\
0
\end{array} \right) \, ,
\end{align}
as well as the CM
\begin{align}
\left( 
\begin{array}{cc}
\langle \hat q_e^2 \rangle              & \frac{1}{2}\langle \hat q_e\hat p_e+\hat p_e\hat q_e \rangle \\
\frac{1}{2}\langle \hat p_e\hat q_e+\hat q_e\hat p_e \rangle & \langle \hat p_e^2 \rangle
\end{array}
\right)
=
\left( \begin{array}{cc}
\frac{1}{1+2n} & 0 \\
0 & 1+2n
\end{array}
\right) \, .
\end{align}


In our QRNG the continuous variable $X$ is mapped into a discrete and bounded variable
$\bar X$ due to the use of an analog-to-digital converter (ADC).
We therefore consider a model in which $X$ is replaced by
a discrete variable $\bar X$ such that
\begin{equation}
p_{\bar{X}}(k) =  
\int_{I_k} dx p_X(x) \, ,
\end{equation}
where $I_k$'s are $d$ intervals that discretize the outcome of homodyne detection.
In a typical setting, these $d$ non-overlapping intervals $I_k$ are of the form 
\begin{align}
I_1 & = ( -\infty , -R] \, , \label{I1} \\ 
I_d & = ( R , \infty ) \, ,  \label{Id}
\end{align}
and for $k=2,\dots, d-1$
\begin{equation}\label{Ik}
I_k = ( a_k - \Delta x/2 , a_k + \Delta x/2 ] \, ,
\end{equation}
with $a_k = - R + (k-1) \Delta x/2$ and $\Delta x = 2R/(d-2)$.
This choice of the intervals reflects the way in which an ideal ADC with range $R$ and bin size $\Delta x$ operates in mapping a continuous variable into a discrete one. However ADCs are not ideal devices, and in the Supplemental Material we show how the digitization error of an ADC reduces the min-entropy.

In terms of the discrete variable $\bar X$, the correlations with the eavesdropper are then described by the state
\begin{equation}\label{thestate}
\rho_{\bar X E} = \sum_{k} p_{\bar{X}}(k) |k\rangle \langle k | \otimes \rho_{E}^{(k)} \, ,
\end{equation}
with
\begin{equation}
\rho_{E}^{(k)} = 
\frac{1}{ p_{\bar{X}}(k) } \int_{I_k} dx p_X(x) \rho_E^x \, .
\end{equation}


We are now ready to quantify the secure rate of the QRNG in terms of the conditional min-entropy.
Given the state $\rho_{\bar X E}$ in Eq.\ (\ref{thestate}), the min-entropy of $\bar X$ conditioned on the eavesdropper (denoted with the letter $E$) reads \cite{Tomamichel2012}
\footnote{Here $\log$ stands for the logarithm in base $2$ and $\ln$
for the natural logarithm.}
\begin{align}
H_\mathrm{min}(\bar X|E)_\rho = \sup_{\gamma} \left[ -\log { \| \gamma_E^{-1/2} \rho_{\bar XE} \, \gamma_E^{-1/2} \|_\infty } \right] \, ,
\end{align}
where $\| \cdot \|_\infty$ denotes the operator norm (equal to the value of the maximum eigenvalue).

Since a direct computation of the min-entropy is not feasible as it requires an optimization over all density operators $\gamma$ in an infinite-dimensional Hilbert space, we instead focus on finding a computable and tight lower bound.
A first lower bound on the min-entropy is obtained by computing
$\| \gamma_E^{-1/2} \rho_{\bar X E} \, \gamma_E^{-1/2} \|_\infty $ for a given choice of the state $\gamma$, so that we have
\begin{align}
H_\mathrm{min}(\bar X|E)_\rho & \geq 
- \log { \| \gamma_E^{-1/2} \rho_{\bar XE} \, \gamma_E^{-1/2} \|_\infty } \\
& = - \log { \left[ \sup_k \, p_{\bar X}(k) \, \| \gamma_E^{-1/2} \rho_{E}^{(k)} \gamma_E^{-1/2} \|_\infty \right] } \, ,
\label{supf}
\end{align}
where the last equality holds because the eigenstates $|k\rangle$ of $\rho_{\bar X E}$ in Eq.\ (\ref{thestate}) are mutually orthogonal.
Here we set $\gamma$ equal to a Gaussian state with zero mean and CM
\begin{align}
\left( \begin{array}{cc}
1+2(n+\delta) & 0 \\
0 & 1+2(n+\delta)
\end{array}
\right) \, ,
\end{align}
where the parameter $\delta$ will be optimized {\it a posteriori} to make the bound as tight as possible.
This choice for the CM is somewhat arbitrary but, as we show in the Supplemental Material, it yields a tight bound on the min-entropy.

A second lower bound is obtained by applying the triangular inequality,
\begin{align}
& p_{\bar X}(k) \, \| \gamma_E^{-1/2} \rho_{E}^{(k)} \gamma_E^{-1/2} \|_\infty \nonumber  \\
& = \| \gamma_E^{-1/2} \int_{I_k} dx \, p_X(x) \, \rho_{E}^x \gamma_E^{-1/2} \|_\infty \\
& \leq \int_{I_k} dx \, p_X(x) \, \| \gamma_E^{-1/2} \rho_{E}^x \, \gamma_E^{-1/2} \|_\infty \, ,
\label{supf1}
\end{align}
which implies
\begin{equation}
H_\mathrm{min}(\bar X|E) \geq 
- \log { \left[ \sup_k  \int_{I_k} \hspace{-0.2cm} dx \, p_X(x) \, \| \gamma_E^{-1/2} \rho_{E}^x \, \gamma_E^{-1/2} \|_\infty
\right] } \, .
\label{supf2}
\end{equation}

Since $\rho_E^{x}$ and $\gamma_E$ are both Gaussian states, the above lower bound can be computed using the Gibbs-representation techniques developed in Ref.~\cite{Banchi2015}.
Employing these techniques and additional tools, Ref.~\cite{Seshadreesan2018} derived a formula for the min-entropy.
By applying this result we obtain \footnote{Note that Theorem 24 of Ref.~\cite{Seshadreesan2018} gives a formula to compute the norm of the operator $\rho^{1/2} \gamma^{-1} \rho^{1/2}$. However, it is easy to show that this is the same as the norm of $\gamma^{-1/2} \rho \gamma^{-1/2}$. 
In fact, consider the operator $\rho^{1/2} \gamma^{-1/2}$, and its singular value decomposition $\rho^{1/2} \gamma^{-1/2} = U \Delta V$, where $U$ and $V$ are unitary and $\Delta$ is diagonal. We then have $\rho^{1/2} \gamma^{-1} \rho^{1/2} = U \Delta^2 U^\dag$.
Analogously, we also have $\gamma^{-1/2} \rho \gamma^{-1/2} = V^\dag \Delta^2 V$.
As the operators $\rho^{1/2} \gamma^{-1} \rho^{1/2}$ and $\gamma^{-1/2} \rho \gamma^{-1/2}$ are unitary equivalent, they also have the same norm.}
\begin{widetext}
\begin{equation}
\int_{I_k} dx \, p_X(x) \, \| \gamma_E^{-1/2} \rho_{E}^x  \gamma_E^{-1/2} \|_\infty 
=  \frac{1}{g}  \frac{(n+\delta)(1+n+\delta)}{\sqrt{2\pi\delta(2 n (n+1+\delta) + \delta)}} 
\int_{I_k} dx \, \exp{ \left[ \frac{-x^2}{2g^2} \frac{\delta}{2n (n+1+\delta)+\delta} 
\right]} \, .
\end{equation}
\end{widetext}

The supremum over $k$ can be computed for any given collection of intervals $I_k$'s.
For intervals as in Eqs.\ (\ref{I1})-(\ref{Ik}) we obtain
\begin{widetext}
\begin{align}
H_\mathrm{min}(\bar X|E) \geq 
-\log{ \left[ \frac{(n+\delta)(1+n+\delta)}{\delta} 
\max\left\{ 
\mathrm{erf}\left( \sqrt{ \frac{\delta}{4n (n+1+\delta)+2\delta} }  \frac{\Delta x}{2g}  \right) ,
\frac{1}{2} \mathrm{erfc}\left( \sqrt{ \frac{\delta}{4n (n+1+\delta)+2\delta} } \frac{R}{g}  \right)
\right\} \right] } \, .
\label{Hmin-final}
\end{align}
\end{widetext}
We remark that this is in fact a family of lower bounds parameterized by $\delta$. 
One should then find the optimal value of $\delta$ for which the bound is tighter.

So far we have considered quantum noise, which was quantified by the mean photon number $n$. If the variance also includes an additive term $\zeta$ due to classical noise, $v^2 = g^2(1+2n) + \zeta^2$, then this can be included in the above analysis by treating it as quantum noise and re-defining $n \to n + \frac{\zeta^2}{2g^2}$.

\subsection{Beyond IID: stationary Gaussian process}\label{Sec:beyondiid}

Going beyond i.i.d., we now consider a more realistic scenario where the measured signal has a finite bandwidth. 
The security analysis of this section holds under the assumption that a potential adversary performs a Gaussian attack. This attack is not necessarily i.i.d. (unlike the collective Gaussian attack considered in the previous section), but it is assumed to be stationary, i.e., it induces a stationary Gaussian process. The assumption of stationarity is thereby in accordance with the device-dependent nature of our scheme.

In this section we first build an i.i.d.\ model for the non-i.i.d.\ process.
Then we apply the results of Section \ref{Sec:iidlimit} to compute a lower bound on the min-entropy with quantum side information.

The analysis deals with two processes. One is the signal $X$, i.e.\ the homodyne measurement of the quantum state including all additive noise processes. The second is the excess noise $U$, i.e.\ all noise sources present in the measurement apart from the pure vacuum fluctuations, for instance electronic noise of the detector and intensity noise of the local oscillator laser. 
When a measurement is performed at a given time $t$, the measured signal is denoted $X_t$.
Similarly, we denote as $U_t$ the excess noise at time $t$ (which we need as a theoretical tool, even though this quantity is not accessible experimentally).
We assume that both $X$ and $U$ are stationary Gaussian processes.

Let us first consider the output $X$ of ideal homodyne detection, and denote as $\sigma^2$ its variance. From the power spectrum $f_X(\lambda)$ we can estimate the entropy rate of the signal \cite{Gray2006}
\begin{align}
h(X) = \frac{1}{2} \log{( 2 \pi e \sigma_X^2 )} \, ,    
\end{align}
where
\begin{align}
\sigma_X^2 = \frac{1}{2 \pi e} 2^{ \int_0^{2\pi} \frac{d\lambda}{2\pi} \log[2\pi e f_X(\lambda)]}    
\end{align}
is the conditional variance.
Similarly, from the power spectrum of the excess noise $f_U(\lambda)$ we obtain the entropy rate of the noise 
\begin{align}
h(U) = \frac{1}{2} \log{( 2 \pi e \sigma_U^2 )} \, ,
\end{align}
where 
\begin{align}
\sigma_U^2 = \frac{1}{2 \pi e} 2^{ \int_0^{2\pi} \frac{d\lambda}{2\pi} \log[2\pi e f_U(\lambda)]}    
\end{align}
is the conditional variance of the excess noise.

Because of the finite bandwidth of the measuring apparatus, both the signal $X_t$ and excess noise $U_t$ at a given time $t$ are correlated with their values at previous times.
To filter out the effects of these correlations we consider the probability density distribution 
of $X_t$ conditioned on all past signal values, 
\begin{align}
p_{X_t}(x_t|x_{<t}) = G( x_t ; \mu_t , \sigma_X^2 ) \, ,
\end{align}
where $x_t$ denotes the possible values of the variable $X_t$ at time $t$, $x_{<t}$ denotes the collection of values of all signals at times $t'<t$, and $\mu_t$ is the mean value.
Note that $\mu_t$ depends on $x_{<t}$, but the conditional variance $\sigma_X^2$ does not depend on time.
This description is consistent with the assumption of a stationary Gaussian process \cite{Gray2006}.
By definition, the conditioned variable at time $t$ is independent on previous signal values. Therefore, we can formally describe it as the outcome of a measurement applied on an i.i.d.\ quantum state with the same variance.
We identify (using the notation of the Section \ref{Sec:iidlimit}):
\begin{align}
\sigma_X^2 \equiv g^2(1+2n) \, .
\end{align} 
We can then write the (unconditional) signal variance $\sigma^2$ as
\begin{align} \label{model_eq1}
\sigma^2 = g^2(1+2n) + \zeta \, ,
\end{align}
where $\zeta = \sigma^2 - \sigma_X^2$,
and the term $\zeta$ is interpreted as classical noise due to the fluctuations of the mean value $\mu_t$, whose variance is $\zeta$.

In summary, we have defined an effective i.i.d.\ model for the non-i.i.d.\ signal.
This i.i.d.\ model has been obtained using the notion of conditional variance.
The i.i.d.\ model is characterized by the parameters $n$ and $g$. 
Therefore, we need a second equation in addition to Eq.\ (\ref{model_eq1}) to determine the model parameters $n$ and $g$ as function of the measured quantities $\sigma^2$, $\sigma_X^2$, and $\sigma_U^2$. This is obtained through the conditional variance of the excess noise.

For the excess noise $U_t$, we can similarly write the probability density distribution conditioned on past values, i.e., 
\begin{align}
p_{U_t}(u_t|u_{<t}) = G( u_t ; \nu_t , \sigma_U^2 ) \, ,
\end{align}
where $u_t$ denotes the possible values of the variable $U_t$ at time $t$, $u_{<t}$ denotes its past values, and $\nu_t$ is the mean value of $U_t$.
The quantity of interest is the conditional excess noise variance $\sigma_U^2$, 
which represents the variance of the excess noise that is virtually independent of previous noise values. 
We identify this variance with the variance of the excess noise in the i.i.d. model of Section \ref{Sec:iidlimit}: 
\begin{align}\label{model_eq2}
\sigma_U^2 \equiv 2 g^2 n \, .    
\end{align}

Endowed with Eq.\ (\ref{model_eq1}) and (\ref{model_eq2}), we are now in the position of determining the parameters $n$ and $g$ of the i.i.d.\ model of the non-i.i.d.\ process. 
Solving for $n$ and $g$ the coupled equations,
\begin{align}
\sigma_X^2 & = g^2 (1+2n) \, , \\
\sigma^2_U & = 2 g^2 n \, ,
\end{align}
we obtain
\begin{align}
g & = \sqrt{ \sigma_X^2 - \sigma_U^2 } \, , \label{g-model}\\
n & = \frac{1}{2}\frac{\sigma_U^2}{\sigma_X^2 - \sigma_U^2} \, .
\end{align}

Finally, we need to account for the classical noise variance $\zeta$. As discussed in Section \ref{Sec:iidlimit}, we incorporate it in the quantum noise by re-defining 
\begin{align}
n & \to n + \frac{\zeta}{2g^2} \\
  & = \frac{1}{2}\frac{\sigma_U^2}{\sigma_X^2 - \sigma_U^2} 
+ \frac{1}{2}\frac{\sigma^2-\sigma^2_X}{\sigma_X^2 - \sigma_U^2} \\
& = \frac{1}{2}\frac{\sigma^2}{\sigma_X^2 - \sigma_U^2}
- \frac{1}{2} \, . \label{n-final}
\end{align}
 
\begin{figure}
    \centering
    \includegraphics{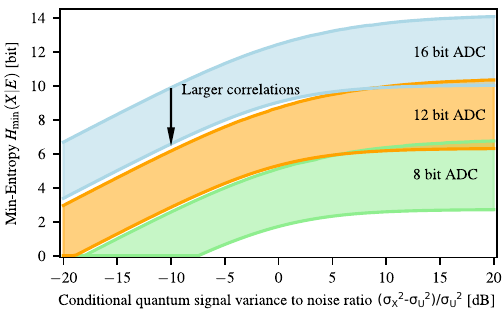}
    \caption{Min-entropy for 8, 12 and 16 bit ADC resolution versus the ratio of conditional variance of the vacuum fluctuations and the conditional variance of the excess noise, $(\sigma_X^2-\sigma_U^2)/\sigma_U^2$. The shaded areas indicate the regions between low correlations ($\sigma_X^2 / \sigma^2 = 0.99$), upper trace, and high correlations ($\sigma_X^2 / \sigma^2 = 0.1$), lower trace. The signal variance has been optimized to obtain the highest min-entropy.}
    \label{fig:theory}
\end{figure}

In conclusion, this model allows us to compute a lower bound for the min-entropy of the non-i.i.d.\ process by inserting the above values for $g$ [in Eq.\ (\ref{g-model})] and $n$ [in Eq.\ (\ref{n-final})] in the min-entropy formula of Eq.\ (\ref{Hmin-final}). 
This is plotted in Fig.~\ref{fig:theory} for varying excess noise, ADC resolution and temporal correlations. 
The x-axis of the plot is the ratio of the conditional variance of the vacuum fluctuations and the excess noise, i.e.\ the quantum noise to excess noise ratio of the virtual i.i.d.\ process. Assuming that the measured homodyne signal and the excess noise have similar temporal correlations, this ratio is independent of the amount of correlations. The amount of correlations present in the system is instead characterized by the ratio $\sigma_X^2/\sigma^2$ which takes the value of 1 for an i.i.d.\ process and becomes smaller for increasing temporal correlations.  
For each ADC resolution the upper traces in the figure show the extractable min-entropy when almost no correlations are present. Obviously, stronger correlations yield lower randomness.

Similar to the result for classical side-information~\cite{Haw2015}, we show that random numbers can be generated for noise treated as quantum side-information as well; and even in the large excess noise regime.  This is due to the fact that relatively small vacuum fluctuations can give a substantial contribution to the entropy if the ADC resolution is sufficiently high. This property is preserved even when a large amount of temporal correlations is present in the recorded data (lower traces).

\begin{figure*}[ht]
  \includegraphics{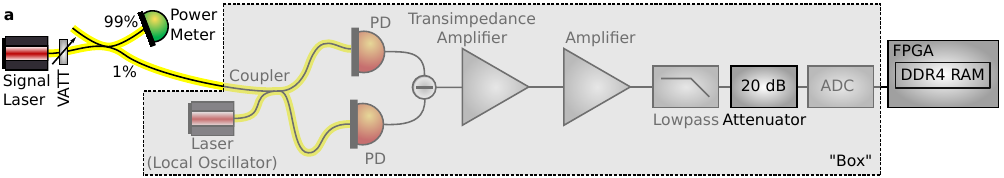} \\
  \includegraphics{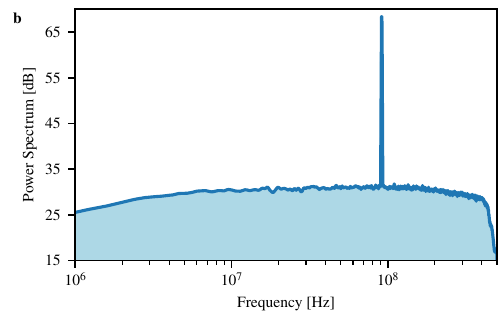}
  \includegraphics{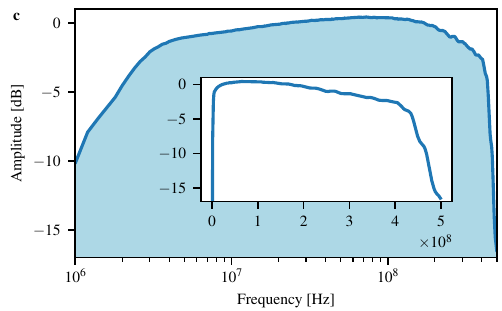}
  \caption{Characterization of the transfer function of the detection system to obtain the vacuum fluctuations noise level. a) Experimental setup for the characterization. b) Power spectrum from a typical measurement. The transfer function is determined by the amplitude of the beat node. c) Transfer function of the homodyne detector and the electronics including the analog-to-digital converter. Inset: Transfer function with linear frequency scale.}
  \label{fig:characterization}
\end{figure*}

\section{System Characterization}

Using the above results, we are now in a position to estimate the min-entropy through a metrology-grade characterization of our setup. According to the security analysis, the min-entropy can be found by determining the variance $\sigma^2$ as well as the conditional variances of the homodyne signal $\sigma^2_X$ and the excess noise $\sigma^2_U$. To obtain a conservative, and thus reliable, estimate of the min-entropy, it is important that the determination of these parameters does not rely on any ideality assumptions of the homodyne detector. In previous studies on homodyne based QRNG, the sole presence of shot noise has been verified by characterizing its scaling with optical power. However, imperfect common-mode rejection, large intensity noise of the laser or stray light coupling into the signal port -- likely to be an issue with integrated photonic chips -- may unnecessarily constrain the extraction of random numbers. Furthermore this characterization method is not directly compatible with metrology-grade characterization as it is difficult to bound the estimation error on the shot noise level with a confidence interval. To circumvent these assumptions and issues we perform an independent, reliable and metrology-grade characterization of the measuring device.

We basically consider the homodyne detector as a black box with an input and an output and minimal assumptions on its internal workings (see Fig~\ref{fig:characterization}a). Our strategy is thus to measure the transfer function of the box and to use this result to conservatively calibrate the power spectral density (PSD) of the vacuum fluctuations. This conservatively estimated result is then compared to the PSD of an actual noise measurement from which we deduce the conditional variances of the signal noise and the excess noise, and finally the min-entropy.

The transfer function of the box is measured by injecting a coherent state in form of a second laser beam (independent of the local oscillator laser) with low power $P_\text{sig}$ into the signal port of the beam splitter as displayed in Fig.~\ref{fig:characterization}a. A typical beat signal is shown in Fig.~\ref{fig:characterization}b obtained by computing an averaged periodogram from the sampled signal. We record the transfer function $\text{TF}(\nu)$ by scanning the frequency of the signal laser. At each difference frequency $\nu$ we determine the power of the beat signal and normalize it to $P_\text{sig}$. At high signal-to-noise ratio the root-mean-square power of the beat signal is purely a function of the coherent state amplitude (i.e.\ the signal laser power) and independent of the noise properties of the two lasers and the detector. 

The transfer function includes the efficiency of the interference, optical loss and the quantum efficiency of the photodiodes, as well as the frequency dependent gain of all amplifiers, the lowpass filter, and the analog bandwidth of the ADC. Since the vacuum noise was amplified to optimally fill the range of the ADC, we used a 20\,dB electrical attenuator with flat attenuation over the frequency band of interest to avoid saturation, see Fig.~\ref{fig:characterization}a. The result of the transfer function characterization, normalized to a maximum gain of 1 is shown in Fig.~\ref{fig:characterization}c. Assuming linearity of the detector we obtain the PSD of the vacuum fluctuations by multiplying the transfer function $\text{TF}(\nu)$ with the shot noise energy $\hbar\omega_L$ contained in 1 Hz bandwidth, where $\hbar$ is Planck's constant and $\omega_L$ is the angular frequency of the local oscillator laser. By modelling the inner workings of the black box, we confirm in the Supplemental Material that with this procedure we indeed obtain a lower bound on the PSD of the vacuum fluctuations.

\begin{figure*}
    \centering
    \includegraphics{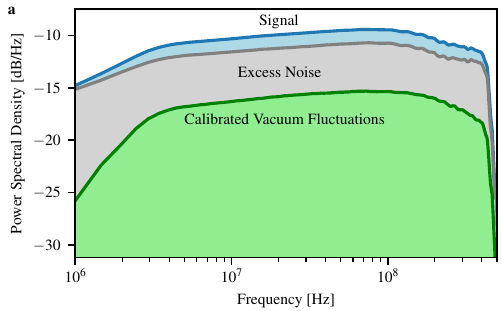}
    \includegraphics{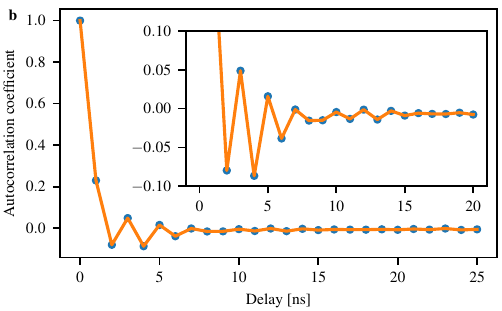} \\
    \includegraphics{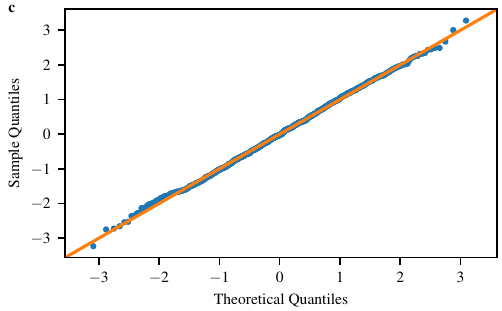}
    \includegraphics{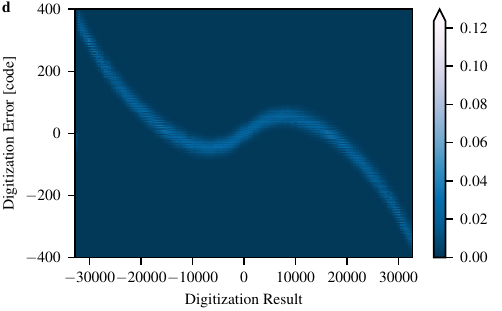}
    \caption{Experimental results. a) The figure shows the power spectral densities of the signal, the calibrated vacuum fluctuations (obtained by the characterization) and the excess noise (obtained by subtracting the PSD of the vacuum fluctuations from the PSD of the signal. b) Autocorrelation coefficients calculated from the measured samples and averaged 1000 times. The inset shows a zoom. c) Q-Q plot indicating the Gaussianity of the measured samples. The variance of the samples has been normalized to 1. The limited ADC range truncates the tails of the Gaussian distribution which results in slight deviations from the theoretical quantiles towards the ends. d) Digitization error of the ADC with respect to the digitization results. The non-linearity and digitization noise of the ADC leads to a reduction of the min-entropy.}
    \label{fig:results}
\end{figure*}

The conservatively estimated PSD of the vacuum fluctuations is shown in Fig.~\ref{fig:results}a together with the actually measured PSD of the signal. 
The spectra are clearly ``colored'' which indicates that the data samples are correlated and therefore non-i.i.d. This is further corroborated in Fig.~\ref{fig:results}b, where the autocorrelation of the signal is plotted. It justifies the importance of using the min-entropy relation associated with non-i.i.d.\ samples.

From the PSDs we calculate the three parameters for obtaining the min-entropy which are summarized in Table~\ref{tab:expresults}. By minimizing the min-entropy over the confidence set of the estimated parameters, we obtain 10.74\,bit per 16 bit sample with a failure probability 
of $\epsilon_\text{PE} = 10^{-10}$ (i.e.\ the probability that the actual parameters are outside the confidence intervals). To verify the Gaussian assumption in our security proof, we calculated the probability quantiles of the measured samples and compared those to the theoretical quantiles of a Gaussian distribution, see Fig.~\ref{fig:results}c.

Finally, we characterized the digitization error of our ADC which is shown in Fig.~\ref{fig:results}d. The measurement protocol is described in the Supplemental Material. Using confidence intervals, the worst case estimate of the reduction of the min-entropy due to the digitization error is 1.77\,bit, thus yielding a total min-entropy of 8.97\,bit. This relatively large reduction is due to the fact that our ADC is 4-way interleaved.

\begin{table}
    \centering
    \begin{tabular}{p{4.1cm}|c|>{\centering\arraybackslash}p{2cm}}
        Parameter & Mean & Confidence interval \\
        \hline
         $\sigma^2$ & $3.96 \times 10^7$ & $0.09 \times 10^7$ \\
         $\sigma_X^2$ & $3.29 \times 10^7$ & $0.07 \times 10^7$ \\
         $\sigma_U^2$ & $2.49 \times 10^7$ & $0.06 \times 10^7$ \\
         Conditional quantum to excess noise ratio & -4.9 dB &\\
         Temporal correlations $\sigma_X^2/\sigma^2$& 0.83 & \\ 
         \hline\hline
         &&\\
         Min-entropy & 10.74 bit & \\
         Reduction due to ADC digitization error & 1.77 bit \\
         Calculated secure length & 2079 bit & \\
         Extracted length & 2048 bit &
    \end{tabular}
    \caption{Summary of parameters determined by the metrological characterization with their confidence intervals for $\epsilon_\text{PE} = 10^{-10}$: signal variance $\sigma^2$, conditional signal variance $\sigma_X^2$ and conditional excess noise variance $\sigma_U^2$. The calculated min-entropy minimized over the confidence intervals, the secure length according to the leftover hash lemma and the length of the extracted random sequence in the experiment.}
    \label{tab:expresults}
\end{table}

\section{Real-Time Randomness Extraction}

Having calculated the min-entropy, the next step is to extract random numbers. This is done by using a strong extractor based on a Toeplitz matrix hashing algorithm in which the seed can be reused~\cite{Wegman1981}.  
We chose matrix dimensions of $n = 4096$ bits and $m = 2048$ bits, which corresponds to 256 input samples with a depth of 16 bit and an output length $m < l$, chosen such that Eq.~(\ref{eqn:securelength}) was fulfilled with $H_\text{min} = 8.97$\,bit and $\epsilon_\text{hash} < 10^{-33}$. The 16\,bit samples provided by the ADC at a rate of 1\,GHz are received by the FPGA in chunks of 64\,bits at a rate of 250\,MHz. For the algorithm implementing the Toeplitz hashing we followed the approach of Ref.~\cite{Zheng2019}. Every clock cycle the 64\,bits were stored in a block until $n$-bits were accepted, after which the next block started receiving data. For each full block, we carried out the hashing multiplication with bit-wise AND and subsequent XOR operations on the Toeplitz matrix by first splitting up the matrix into submatrices of width 16\,bit, and then shifting the data through the operations. When the hashing was completed, the m-bit wide output data was stored in a register, and the next block was processed. The achieved throughput was 8 Gbit/s.

\section{Conclusion}

In conclusion, we have demonstrated a QRNG based on the measurement of vacuum fluctuations with a real-time extraction at a rate of 8\,GBit/s. Our QRNG has a strong security guarantee with a failure probability of $N^\prime \cdot \epsilon_\text{hash} + \epsilon_\text{PE} + \epsilon_\text{seed} = N^\prime \cdot 10^{-33} + 10^{-10} + \epsilon_\text{seed}$, where $N^\prime$ is the number of QRNG runs in the past, $\epsilon_\text{hash}$ is the security parameter related to the removal of side information (see Eq.~\ref{eqn:securelength}), $\epsilon_\text{PE} = 10^{-10}$ is the security parameter of the metrological grade parameter estimation and $\epsilon_\text{seed}$ describes the security of the random bits used for seeding the randomness extractor. Since an adversary may have access to all quantum side information from the past, $\epsilon_\text{hash}$ grows with time~\cite{Frauchiger2013}. We chose a value of $10^{-33}$ to be able to generate Gaussian random numbers with security $\epsilon = 10^{-9}$ for a single execution of a continuous variable quantum key distribution (QKD)~\cite{Pirandola2019} protocol with $10^{10}$ transmitted quantum states even after 10 years of continuous operation of the QRNG. See the Supplemental Material for details.
In our experiment the seed bits were chosen with a pseudo-random number generator, which did not allow us to give a security guarantee for $\epsilon_\text{seed}$. The generated random numbers passed both the dieharder~\cite{Brown2018} and the NIST~\cite{Rukhin2001} statistical batteries of randomness tests.

Due to the choice of a very small $\epsilon_\text{hash}$, the real-time speed of our QRNG was limited to 8 GBit/s by the input size of the Toeplitz extractor required by our FPGA implementation. Without limitations to the matrix size a speed of 8.9 GBit/s could be reached. The main limitations are the very conservative estimates of the min-entropy reduction due to the ADC digitization error and the shot noise calibration.

Nevertheless, our QRNG is perfectly suited for use in high-speed QKD links, for instance in GHz clocked discrete variable~\cite{Dixon2008} as well as in high-speed continuous-variable QKD~\cite{Huang2015b}. For Gaussian-modulated CVQKD the uniform random number distribution has to be converted to a Gaussian distribution which requires a larger random number generation rate. Furthermore, QKD requires composable security and a guarantee of privacy of the random numbers as provided by our system. 

Further developments to guarantee reliable operation over a long time and to fulfill requirements by certification authorities would need to include power-on self-tests and online testing of the parameters in the security proof as well as the generated random numbers.

\section*{Acknowledgements}
The authors acknowledge support from the Innovation Fund Denmark through the Quantum Innovation Center, Qubiz. TG, AK, DSN, NJ and ULA acknowledge support from the Danish National Research Foundation, Center for Macroscopic Quantum States (bigQ, DNRF142). TG, NJ, SP and ULA acknowledge the EU project CiViQ (grant agreement no.\ 820466).  The authors thank Alberto Nannarelli for valuable discussions.

\onecolumngrid
\appendix

\section{ADC Characterization}
\subsection{Min-entropy correction due to digitization error}
Consider a model of ADC noise where the state
\begin{equation}\label{thestate0}
\rho_{\bar X E} = \sum_{k} p_{\bar{X}}(k) |k\rangle \langle k | \otimes \rho_{E}^{(k)} \, ,
\end{equation}
is replaced by its noisy version
\begin{equation}\label{thestate1}
\rho_{\bar X' \bar X E} = \sum_{k,k'} p_{\bar X'|\bar X}(k'|k) p_{\bar{X}}(k) 
|k'\rangle \langle k' | \otimes |k\rangle \langle k | \otimes \rho_{E}^{(k)} \, .
\end{equation}

Consider for now a specific measurement that maps the quantum state on $E$ into a random variable $Z$. Therefore, we can write the joint probability distribution of $\bar X'$, $\bar X$, $\bar Z$:
\begin{align}
p_{\bar X' \bar X  Z}(k',k,z)
& = p_{\bar X'|\bar X}(k'|k)  
p_{Z|\bar{X}}(z|k) p_{\bar{X}}(k)\\
& = p_{\bar X'|\bar X}(k'|k) 
p_{\bar{X}|Z}(k|z) p_{Z}(z) \, ,
\end{align}
where the second equality follows from the Bayes rule.

Note that the variable $\bar X$ is not observed by the user. Furthermore, we assume that it is not observed nor controlled by the adversary.
This means that we are interested, for a given measurement performed by the adversary, on the conditional probability:
$$
p_{\bar X' | Z}(k'|z)
= \sum_k p_{\bar X'|\bar X}(k'|k) 
p_{\bar{X}|Z}(k|z) \, .
$$

The following holds:
\begin{align}
p_{\bar X' | Z}(k'|z)
& = \sum_k p_{\bar X'|\bar X}(k'|k) 
p_{\bar{X}|Z}(k|z) \\
& \leq \sum_k p_{\bar X'|\bar X}(k'|k) 
\max_h p_{\bar{X}|Z}(h|z) \, .
\end{align}

From this, we obtain an upper bound on the guessing probability:
\begin{align}
P_\mathrm{guess}({\bar X'}|Z)
& = \sum_z \max_{k'} p_{\bar X' | Z}(k'|z) p_Z(z) \\
& \leq \max_{k'} \sum_k p_{\bar X'|\bar X}(k'|k) 
\sum_z \max_h p_{\bar{X}|Z}(h|z) p_Z(z) \\
& = \max_{k'} \sum_k p_{\bar X'|\bar X}(k'|k) 
P_\mathrm{guess}({\bar X}|Z) \, .
\end{align}

The above inequality holds for any measurement $E \to Z$.
Therefore it also holds for the optimal measurement.
This implies a bound for the probability of guessing with quantum side information:
\begin{align}
P_\mathrm{guess}({\bar X'}|E)
& \leq \max_{k'} \sum_k p_{\bar X'|\bar X}(k'|k) 
P_\mathrm{guess}({\bar X}|E) \, .
\end{align}

In terms of the min-entropy, this reads
\begin{align}\label{min-entropy-c1}
H_\mathrm{min}({\bar X'}|E) 
& \geq H_\mathrm{min}({\bar X}|E)
- \log{\left[ \max_{k'} \sum_k p_{\bar X'|\bar X}(k'|k) \right]} 
 \, .
\end{align}

\subsection{Estimation of the min-entropy correction}

Note that the quantity in the argument of the logarithm in Eq.\ (\ref{min-entropy-c1}) is proportional to a probability:
\begin{align}
\max_{k'} \sum_k p_{\bar X'|\bar X}(k'|k) 
= N \max_{k'} \sum_k p_{\bar X'|\bar X}(k'|k) \frac{1}{N}  
= N \max_{k'} p_{\bar X'}(k') \, ,
\end{align}
where $N$ is the cardinality of $\bar X'$ and $\bar X$, and $p_{\bar X'}(k') = \sum_k p_{\bar X'|\bar X}(k'|k) \omega_{\bar X}(k)$, where $\omega_{\bar X}(k) = \frac{1}{N}$ is the flat distribution for ${\bar X}$.

The probability $p_{\max} := \max_{k'} p_{\bar X'}(k')$ can be estimated experimentally. 
If using a sample of size $S$ the most common value of $k'$ is obtained $S_{\max}$ times, then our best guess for $p_{\max}$ is the relative frequency $\nu_{\max} = \frac{S_{\max}}{S}$. 
A confidence interval can then be obtained from the Hoeffding inequality:
\begin{align}
    \mathrm{Pr}\left\{ p_{\max} \geq \nu_{\max} + \delta \right\} \leq e^{-2  \delta^2 S} \, ,
\end{align}
that is,
\begin{align}
    \mathrm{Pr}\left\{ p_{\max} \geq \nu_{\max} + \sqrt{\frac{1}{2S}\ln{\frac{1}{\epsilon}}} \right\} \leq \epsilon \, .
\end{align}

In conclusions, we obtain that the following min-entropy bound holds up to a probability smaller than $\epsilon$:
\begin{align}
H_\mathrm{min}({\bar X'}|E) 
& \geq H_\mathrm{min}({\bar X}|E)
- \log{\left( N \nu_{\max} + N \sqrt{ \frac{1}{2S} \ln{\frac{1}{\epsilon}}} 
\right)} 
 \, .
\end{align}

For the experimental characterisation of the ADC, we have used a sample of size $S = 2^{18} \times 500000$, and the cardinality was $N=2^{16}$. The relative frequency of the maximum value was such that $N \nu_{\max} \simeq 2.79$.
Putting $\epsilon = 10^{-10}$, we then obtain the correction term
\begin{align}
    N \sqrt{ \frac{1}{2S} \ln{\frac{1}{\epsilon}}} = 0.128 \sqrt{\ln{\frac{1}{\epsilon}}} = 0.128 \sqrt{10\ln{10}} \simeq 0.6 \, .
\end{align}

\subsection{Measurement}

\begin{figure}
    \centering
    \includegraphics{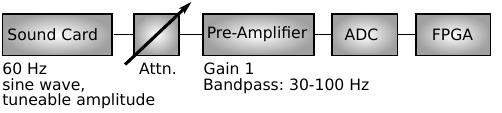}
    \caption{Measurement setup for ADC characterization. The sound card has a high impedance output which was converted to 50 Ohm by the pre-amplifier.}
    \label{fig:setup_adc_characterization}
\end{figure}

The setup for the ADC characterization is shown in \ref{fig:setup_adc_characterization}. A sine wave of 60 Hz frequency is generated by a 24 bit sound card. The maximum amplitude of the sine is adjusted by a variable attenuator to match the input range of the ADC. A pre-amplifier (Standard Research Systems SR560) is used to bandpass filter the signal and as buffer to convert the high impedance output of the sound card to a 50 Ohm impedance as required by the ADC.

The measurement was performed as follows: We swept the amplitude of the sine wave from 0 to $2^{23}-1$ in steps of 64. This yields 4 measurements with different voltage levels for each bin of the 16 bit ADC. For each amplitude setting 10 periods of the sine wave were recorded by the ADC with a sampling rate of 1 GS/s and the data was transferred to a computer via the FPGA. We analysed the data by determining the location of the maxima and minima and for each we calculated a histogram from 50,000 samples around the maximum and minimum, respectively. The 10 histograms were summed so that we obtained a histogram from 500,000 samples. To obtain $p_{\bar X'|\bar X}(k'|k)$ we combined the 4 measurements per ADC bin and normalized the probability distribution.

\section{Comparison with classical side information}\label{HomoUB}

In this Section we compare our lower bound on the min-entropy with quantum side information with an upper bound obtained under the assumption that the eavesdropper performs ideal homodyne detection.

In the main body of the paper we have obtained the following lower bound on the min-entropy with quantum side information:
\begin{align}
H_\mathrm{min}(\bar X|E) \geq 
-\log{ \left[ \frac{(n+\delta)(1+n+\delta)}{\delta} 
\max\left\{ 
\mathrm{erf}\left( \sqrt{ \frac{\delta}{4n (n+1+\delta)+2\delta} }  \frac{\Delta x}{2g}  \right) ,
\frac{1}{2} \mathrm{erfc}\left( \sqrt{ \frac{\delta}{4n (n+1+\delta)+2\delta} } \frac{R}{g}  \right)
\right\} \right] } \, .
\label{Hmin-final-app}
\end{align}

For the sake of comparison, we simplify this expression by using an optimal choice for the ADC range $R$. This yields \begin{equation}\label{LB3}
H_\mathrm{min}(\bar X|E)_\rho \geq 
-\log{ \left[
\frac{(n+\delta)(1+n+\delta)}{\delta}  
\, \mathrm{erf}\left( \sqrt{ \frac{\delta}{4n (n+1+\delta)+2\delta} }  \frac{\Delta x}{2g} \right) 
\right] } \, .
\end{equation}

At the lowest order in $\Delta x$ this in turn becomes 
\begin{equation}\label{LB4}
H_\mathrm{min}(\bar X|E)_\rho \gtrsim
-\log{\left( \frac{\Delta x}{g} \right)} 
-\log{ \left( \frac{(n+\delta)(1+n+\delta)}{\sqrt{2\pi \delta 2n (n+1+\delta)+\delta]}} \right)}  \, .
\end{equation}

This bound can be made tighter by using an optimal value for $\delta$. For example, putting $\delta = n$ we obtain
\begin{equation}\label{lowbound_comp}
H_\mathrm{min}(\bar X|E)_\rho \gtrsim 
-\log{\left( \frac{\Delta x}{g} \right)} 
-\log{ \left( \frac{\sqrt{2}(2n+1)}{\sqrt{\pi(4n+3)}} \right)}  \, .
\end{equation}

\vspace{0.5cm}

Let us now compute an upper bound for an eavesdropper measuring by homodyne detection. If user and eavesdropper both apply homodyne detection, then they generate a pair of correlated Gaussian variables $X$ and $Y$ such that
\begin{align}
p_Y(y) & = G( y ; 0 , 1+2n) \, ,\\
p_X(x|y) & = G\left( x ; \frac{2g\sqrt{n(n+1)}}{1+2n} y , \frac{g^2}{1+2n} \right) \, .
\end{align}

The variable $X$ is then mapped into a discrete and bounded variable $\bar X$
as described in the main body of the paper.
Using an optimal choice of the range $R$ of the ADC, the min-entropy of $\bar X$ conditioned on the $Y$ is
\begin{equation}\label{Homo0}
H_\mathrm{min}(\bar X|C) = - \log{\mathrm{erf}{\left( \frac{\Delta x}{g}\frac{\sqrt{2+4n}}{4} \right) } } \, ,
\end{equation}
or, up to correction of order higher than $\Delta x$, 
\begin{equation}\label{Homo1}
H_\mathrm{min}(\hat X|C) \simeq - \log{\left( \frac{\Delta x}{g} \right)}
- \log{\left( \sqrt{ \frac{1+2n}{2\pi } } \right)} \, .
\end{equation}

\begin{figure}[hb]
\centering
\includegraphics[width=0.4\textwidth]{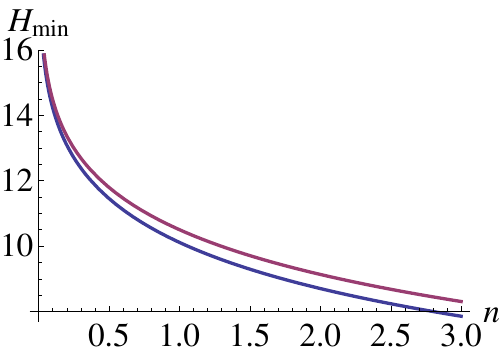}
\caption{Homodyne upper bound $H_\mathrm{min}(\bar X|C)$ as in Eq.\ (\ref{Homo0}) [red line] 
and the min-entropy lower bound in Eq.\ (\ref{lowbound_comp}), vs the mean photon number $n$, for $\delta = n$ and $\Delta x = n/1000$.
} \label{Hmin}
\end{figure}

\begin{figure}[hb]
\centering
\includegraphics[width=0.4\textwidth]{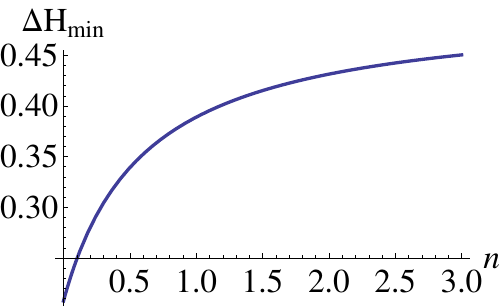}
\caption{The difference between the Homodyne upper bound in Eq.\ (\ref{Homo1}) and the min-entropy lower bound in Eq.\ (\ref{lowbound_comp}), vs the mean photon number $n$, for $\delta = n$.
} \label{HDiff}
\end{figure}

Figure \ref{Hmin} shows, for $\Delta x = n/1000$, the homodyne upper bound in Eq.\ (\ref{Homo0}) and the min-entropy lower bound in Eq.\ (\ref{lowbound_comp}), as function of the mean photon number $n$. The latter is plotted for $\delta = n$.
Figure \ref{HDiff} shows instead the difference between the 
lower bound in Eq.\ (\ref{lowbound_comp}) and the homodyne upper bound in Eq.\ (\ref{Homo1}), again for $\delta = n$
(notice that the difference is independent of $\Delta x$).
This plots whow that our lower bound is tight and in fact homodyne is close to be the optimal measurement for Eve. Note that the difference between the upper and lower bounds is as small as a fraction of a bit.

\section{Estimation of variances and the entropy rate}

In this appendix we discuss the estimation of the variance, entropy rate and conditional
variance of the noise and signal. To make things more concrete, we focus on the
estimation of the signal variance $\sigma^2$, entropy rate $h(X)$ and the conditional signal variance $\sigma_X^2$.
Assume that $T$ is the runtime of the experiment, and $n$ signal measurements are 
performed at regular time intervals of $\delta t = T/n$.
The spectral density computed from these data is a function of $n$ discrete frequencies,
denoted as $\omega_j$'s, taking values between $2\pi/T$ and $2\pi n /T$. 
Below we work with the discrete variable $\lambda_j$ defined as $\lambda_j \equiv T \omega_j/n$, 
which can be approximated by the continuous variable $\lambda$ taking values with domain
$[0,2\pi]$.

We estimate the spectral density $f(\lambda)$ by applying the Welch's method, according
to which the data are first divided in $M$ (possibly overlapping) blocks, and then in each
block the periodogram is computed, i.e., the discrete Fourier transform of the data contained
in that very block. The spectral density is then estimated by taking the average over the 
periodograms. 
We assume that the periodograms, as random variables, are independent and identically 
distributed, and that each periodogram is distributed as the square of a Gaussian variable.
Then the Welch's estimate of the spectral density is distributed as a (rescaled) $\chi^2(k)$ variable
with $M$ degrees of freedom. 
Denoting as $f_0(\lambda_j)$ the Welch's estimate for the spectral density and 
as $f(\lambda_j)$ its real value, then we can obtain a confidence interval by applying 
a tail bound of a $\chi^2(k)$ variable. For example we can exploit the tail bounds 
(see e.g.\ \cite{Lupo2018})
\begin{align}
\mathrm{Pr} \left\{  f(\lambda_j) < \frac{f_0(\lambda_j)}{1+t}  \right\} & \leq e^{- M t^2/8} \, , \\
\mathrm{Pr} \left\{  f(\lambda_j) > \frac{f_0(\lambda_j)}{1-t}  \right\} & \leq e^{- M t^2/8} \, .
\end{align}
For $t \ll 1$ this yields, up to higher order terms,
\begin{align}
\mathrm{Pr} \left\{ f(\lambda_j) \not \in [ (1-t)f_0(\lambda_j)  , (1+t)f_0(\lambda_j) ] \right\} = P(t) \, 
\end{align}
with 
\begin{align}
P(t) \leq 2 \, e^{- M t^2/8} \, .
\end{align}


Let us first discuss the estimation of the entropy rate
\begin{align}
h(X) = \frac{1}{2} \int_0^{2\pi} \frac{d\lambda}{2\pi} \log{ [ 2 \pi e f(\lambda) ]} \, ,
\end{align}
as approximated by the finite sum
\begin{align}
h(X) \simeq \frac{1}{2} \sum_{j=1}^n \frac{1}{n} \log{ [ 2 \pi e f(\lambda_j) ]} \, .
\end{align}

For each given $j$, $1-P(t)$ is the probability that 
$f(\lambda_j) \in [ (1-t)f_0(\lambda_j)  , (1+t)f_0(\lambda_j)  ]$, then
it follows (from an application of the union bound) that
\begin{align}
\mathrm{Pr} \left\{ \exists j \, \, | \, \, f(\lambda_j) \not \in [ (1-t)f_0(\lambda_j)  , (1+t)f_0(\lambda_j)  ] \right\} \leq n P(t) \, .
\end{align}
This is equivalent to say that, with probability larger than $1-n P(t)$,
$f(\lambda_j)$ lays between $(1-t)f_0(\lambda_j)$ and $(1+t)f_0(\lambda_j)$
for all $j=1,\dots,n$.
Therefore
\begin{align}
h(X) \in \left[ 
\frac{1}{2} \sum_{j=1}^n \frac{1}{n} \log{ [ 2 \pi e f_0(\lambda_j) ] } + \frac{1}{2} \log{(1-t)}
, 
\frac{1}{2} \sum_{j=1}^n \frac{1}{n} \log{ [ 2 \pi e f_0(\lambda_j) ]} + \frac{1}{2} \log{(1+t)} 
\right]
\end{align}
with probability at least equal to $1-n P(t) = 1 - 2 n e^{- M t^2/8}$.
A further linear approximation for $t \ll 1$ yields the confidence interval
\begin{align}
h(X) \in \left[ 
\frac{1}{2} \sum_{j=1}^n \frac{1}{n} \log{ [ 2 \pi e f_0(\lambda_j) ] } - \frac{\log{e}}{2} \, t
, 
\frac{1}{2} \sum_{j=1}^n \frac{1}{n} \log{ [ 2 \pi e f_0(\lambda_j) ]} + \frac{\log{e}}{2} \, t
\right]
\, .
\end{align}

Finally, to take into account the overlap between adjacent periodograms,
we replace $M \to \gamma M$, for $\gamma<1$. For example, if the periodogram
have a $50\%$ overlap we put $\gamma = 1/2$.
In conclusion, with an overlap of $50\%$, we obtain that for any 
given $\epsilon >0$, the entropy rate lies within the interval 
\begin{align}
h(X) \simeq
\frac{1}{2} \sum_{j=1}^n \frac{1}{n} \log{ [ 2 \pi e f_0(\lambda_j) ] } \pm 2 \log{e} \, \sqrt{ \frac{1}{M} \ln{\left(\frac{2n}{\epsilon}\right)} }
\, ,
\end{align}
up to a probability not larger than $\epsilon$.


From the entropy rate we obtain a confidence interval for
the conditional variance, 
$\sigma_X \in [ \sigma_X^- , \sigma_X^+]$,
where
\begin{equation}
\sigma_X^\pm 
= \frac{1}{2\pi e} 2^{
\sum_{j=1}^n \frac{1}{n} \log{ [ 2 \pi e f_0(\lambda_j) ] } }
\, 
2^{\pm 4 \log{e} \, \sqrt{ \frac{1}{M} \ln{\frac{2n}{\epsilon}} }
} \, .
\end{equation}


Similarly, we obtain an estimate of the signal
variance $\sigma^2$ by exploiting the relation
\begin{equation}
\sigma^2 = \int_0^{2\pi} \frac{d\lambda}{2\pi} f(\lambda) \, ,
\end{equation}
from which we derive a confidence interval
\begin{equation}
\sigma^2 \simeq
\left( 
1 
\pm 
4 \sqrt{\frac{1}{M} \ln{\frac{2n}{\epsilon}}} 
\right) \sum_{j=1}^n \frac{1}{n} f_0(\lambda_j) \, .
\end{equation}


Along the same lines we obtain a confidence interval for the
conditional noise variance, 
$\sigma_U \in [ \sigma_U^- , \sigma_U^+]$
(this must additionally includes systematic errors).
To obtain a worst-case estimate of the min-entropy
we consider the smaller value for the signal variance, 
$\sigma_X^-$, and the larger one for the noise,
$\sigma_U^+$.

\section{Characterization of vacuum fluctuations power spectral density}

Here, we open up the homodyne detector black box and show by including imperfections that the bound given in the main text is indeed a lower bound on the vacuum fluctuations. As described in the main text we beat two lasers, the local oscillator with power $P_\text{LO}$ and an auxiliary signal laser with power $P_\text{sig}$ which is frequency detuned with respect to the local oscillator by $\nu$. The beams interfere at a beam splitter with splitting ratio $R(\nu):1-R(\nu)$, where the frequency dependence $\nu$ accounts for a frequency dependent common mode rejection of the homodyne electronics. We furthermore take into consideration the visibility of the interference $\chi \in (0,1]$ and the quantum efficiencies $\eta_1$ and $\eta_2 \in (0,1]$ of the two photo diodes.

After photo detection and current subtraction the beat signal current at time $t$ reads
\begin{equation}
    i_\text{beat}(t) = 2\chi^2(\eta_1+\eta_2)\sqrt{R(\nu)(1-R(\nu))}\frac{e}{\hbar \omega}\sqrt{P_\text{LO} P_\text{sig}} \cos(2\pi\nu t)\ .
\end{equation}
Here $\omega$ is the absolute angular frequency of the local oscillator laser. The square of the root mean square (RMS) amplitude of the beat signal digitized by an analog-to-digital (ADC) converter as obtained by a power spectrum of acquired samples is then given by
\begin{equation}
    \widetilde{\text{TF}}(\nu) := \left(\sqrt{2}\chi^2(\eta_1+\eta_2)\sqrt{R(\nu)(1-R(\nu))}\frac{e}{\hbar \omega}\right)^2 P_\text{LO} P_\text{sig}G(\nu)\ ,
    \label{eqn:transferfunction}
\end{equation}
where $G(\nu)$ describes the overall gain of homodyne detector, possible filters and ADC analog input as well as includes the digitization into integers. We call $\text{TF} := \widetilde{\text{TF}}/P_\text{sig}$ the transfer function.

The power spectral density (PSD) of the vacuum fluctuations after photo detection and digitization reads
\begin{equation}
    \text{PSD}_\text{vac} = 2e(i_\text{dc1} + i_\text{dc2})G(\nu) = 2 \frac{e^2}{\hbar \omega}\left(\eta_1(1-R(\nu)) + \eta_2 R(\nu)\right) P_\text{LO} G(\nu)\ ,
\end{equation}
where $i_\text{dc1}$ and $i_\text{dc2}$ are the direct photo currents generated by the photo diodes.
Using the characterization of the transfer function from Eq.~(\ref{eqn:transferfunction}) yields
\begin{equation}
    \text{PSD}_\text{vac} = \hbar\omega \frac{1}{\chi^2}\frac{\eta_1(1-R(\nu)) + \eta_2 R(\nu)}{(\eta_1+\eta_2)^2 R(\nu)(1-R(\nu))} \frac{\widetilde{\text{TF}}(\nu)}{P_\text{sig}} \ge \hbar\omega \frac{\widetilde{\text{TF}}(\nu)}{P_\text{sig}}\ .
\end{equation}
In the last step we lower bounded the PSD of the vacuum fluctuations by using $1/\chi \ge 1$ and 
\begin{align}
\left(\eta_1(1-R(\nu)) + \eta_2 R(\nu)\right)/\left((\eta_1+\eta_2)^2 R(\nu)(1-R(\nu))\right) \ge 1 \, ,
\end{align} 
where equality holds for $\eta_1 = \eta_2 = 1$, $R=0.5$.

\section{Continuous-variable quantum key distribution application example}

Continuous-variable quantum key distribution uses a Gaussian modulation of the coherent state excitation, i.e.\ of both the amplitude and phase quadrature components. For a single execution of the protocol a large amount of coherent quantum states, say $10^{10}$ are generated and transmitted. For each quantum state 2 random numbers are required, the amplitude and the phase quadrature values. In practise these are discretized and for our example we choose an 8 bit resolution. Thus, 16 bit are required to generate 1 coherent state. Since our QRNG delivers 2048 random bits per single execution of the randomness extraction, $N = \frac{16 \text{bit} \cdot 10^{10}}{2048 \text{bit}} \approx 7.8 \times 10^7$ runs are needed.

We now assume that the QRNG ran $N_1$ times prior to the generation of the random numbers for QKD. Then the random number string for QKD has an epsilon stemming from the hashing of

\begin{equation*}
\epsilon_\text{hash}^\text{QKD} = (N_1+1)\epsilon_\text{hash} + (N_1+2) \epsilon_\text{hash} + \ldots + (N_1+N) \epsilon_\text{hash} = \left(N \cdot N_1 + \frac{(N\cdot(N+1))}{2}\right)\epsilon_\text{hash}\ .
\end{equation*}

In our example we require that the random numbers used in the quantum key distribution system have an epsilon security parameter of less than $10^{-9}$. Assuming that the parameter estimation epsilon is constant over time and noticing that $\epsilon_\text{PE}$ is one order of magnitude smaller, the series of random numbers must have an epsilon of $\epsilon_\text{hash}^\text{QKD} \le 10^{-9}$ from hashing.

After continuously running the QRNG for 10 years,
\begin{equation*}
    N_1 = \frac{10\,\text{years}}{ \frac{256}{1 \text{GS}/\text{s}}} \approx 1.2\times 10^{15}\ ,
\end{equation*}
where 256 is the number of acquired samples from the ADC per extraction run.

Solving the above equation for $\epsilon_\text{hash}$ and plugging in numbers for $N_1$ and $N$ yields $\epsilon_\text{hash} \lesssim 10^{-32}$. Thus we chose $\epsilon_\text{hash} = 10^{-33}$ to be on the safe side.

\bibliography{literature}

\end{document}